\documentclass[aps,prl,twocolumn,showpacs,groupedaddress]{revtex4}
\usepackage{epsf}
\begin{document}
\preprint{\today}
\title{Effect of measurement probes upon the conductance of an 
interacting nano-system: \\ 
Detection of an attached ring by non local many body effects}

\author{Axel Freyn}
%
\affiliation{Service de Physique de l'\'Etat Condens\'e (CNRS URA 2464), 
DSM/DRECAM/SPEC, CEA Saclay, 91191 Gif-sur-Yvette Cedex, France}
\author{Jean-Louis Pichard}
%
\affiliation{Service de Physique de l'\'Etat Condens\'e (CNRS URA 2464), 
DSM/DRECAM/SPEC, CEA Saclay, 91191 Gif-sur-Yvette Cedex, France}

\begin{abstract} 
We consider a nano-system connected to measurement probes via leads. 
When a magnetic flux is varied through a ring attached to one lead at 
a distance $L_c$ from the nano-system, the effective nano-system 
transmission $|t_s|^2$ exhibits Aharonov-Bohm oscillations if 
the electrons interact inside the nano-system. These oscillations 
can be very large, if $L_c$ is small and if the nano-system has almost 
degenerate levels which are put near the Fermi energy by a local gate. 
\end{abstract}

\pacs{71.10.-w,72.10.-d,73.23.-b}  


\maketitle

%
%

 To map the low temperature behavior of an interacting system onto 
that of an effective non interacting system with renormalized 
parameters is the basis of the simplest many body theories (Hartree-Fock 
(HF) approximation \cite{fetter}, Landau theory of dressed 
quasi-particles). In one dimension (1d), this mapping fails to describe 
the collective excitations of a macroscopic wire (Luttinger-Tomonaga limit), 
but can be used if the electrons interact only inside a microscopic part of 
a very long wire. This corresponds \cite{imry} to the set-up used for 
measuring the quantum conductance $g$ of a nano-system with two attached  
probes. The electrons can strongly interact inside the nano-system 
(molecule, quantum dot with a few electrons, atomic chains created in 
a break junction) while their interaction can be neglected outside. 
If the many body scatterer can be mapped onto an effective one body 
scatterer, detecting the presence of interactions from a zero temperature 
transport measurement looks difficult. Fortunately, the interactions give 
rise to a new phenomenon which does not exist in a bare one body scatterer: 
the effective transmission  $|t_s|^2$ ceases \cite{mwp,afp} to be local. 
We study in this letter a set-up for detecting by a conductance measurement 
the non locality of $|t_s|^2$ due to nano-system interactions.  
\begin{figure}
\centerline{
\epsfxsize=8.5cm 
\epsfysize=5cm 
\epsffile{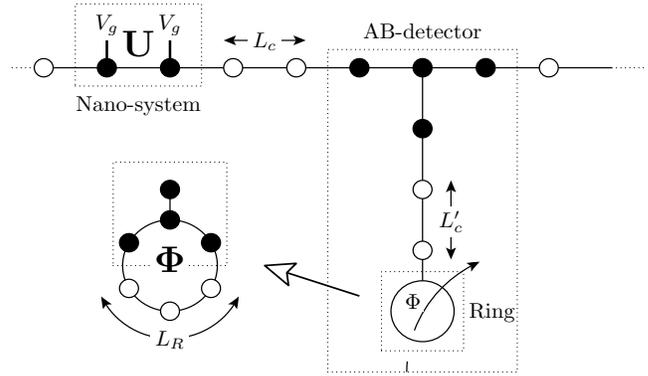}
}
\caption{Considered set-up made of a many body scatterer with two 
semi-infinite 1d leads: Polarized electrons interact only inside the 
nano-system (two sites with inter-site repulsion $U$, hopping term 
$t_d$ and applied gate voltage $V_G$). A ring is attached at a distance 
$L_c$ from the nano-system.} 
\label{fig1} 
\end{figure} 

The idea can be simply explained using the HF approximation \cite{fetter} 
and the Landauer formulation \cite{imry} of quantum transport. Let us assume 
a nano-system with interactions in contact with two 1d non interacting leads. 
Using the HF approximation, one can map the ground state of this set-up onto 
that of an effective one body scatterer with leads, introducing HF 
corrections which probe energy scales much below the Fermi energy and 
length scales much larger than the nano-system size. Via the conduction 
electrons, one can induce Friedel oscillations of the electron density 
inside the nano-system by inserting a second scatterer in one lead at a 
distance $L_c$ from the nano-system. These oscillations change the 
Hartree and Fock terms of the nano-system effective Hamiltonian. This 
generates a non local effect upon its effective transmission $|t_s|^2$, and 
hence upon the Landauer conductance of a set-up embedding the nano-system. 
This non local effect upon  $|t_s|^2$ decays as $1/L_c$ with $\pi/k_{F}$ 
oscillations ($k_{F}$ being the Fermi momentum), if it is driven by Friedel 
oscillations in 1d leads ($1/L_c^d$ decay for d-dimensional leads). This 
phenomenon is reminiscent of the RKKY interaction \cite{kittel} between 
magnetic moments via conduction electrons. At a temperature $T \neq 0$, this 
non local effect is exponentially suppressed \cite{afp} when $L_c$ exceeds 
$L_T$, the scale on which the electrons propagate at the Fermi velocity 
during a time $\propto 1/T$. 

 We use a simple 1d toy model of polarized electrons (spinless 
fermions) for which the Hartree-Fock approximation allows to 
describe \cite{afp} the effect of a repulsion $U$ acting between two 
consecutive sites only. To detect how the nano-system effective 
transmission $|t_s|^2$ depends on the measurement probes, we put a 
second (one body) scatterer in one of the leads at a distance $L_c$ 
from the nano-system.  
Hereafter, we refer to this second scatterer as the AB-detector, 
since it includes a ring threaded by an Aharonov-Bohm (AB) flux $\Phi$, 
the combined set-up being sketched in FIG.\ \ref{fig1}. Flux dependent 
Friedel oscillations of the electron density are induced inside the 
nano-system by the AB-detector. If the electrons interact inside the 
nano-system, $|t_s|^2$ exhibits periodic AB-oscillations, which vanish if 
the ring is too far from the nano-system or if the electrons cease to 
interact. 

%
%
 Our toy model is a tight-binding model on an infinite 1d lattice, where 
spinless fermions do not interact, unless they occupy the nano-system 
(two central sites $0$ and $1$), which costs an energy $U$. A potential 
$V_G$ can be varied inside the nano-system by a local gate. The Hamiltonian 
of the nano-system with the leads reads:
\begin{equation}
H = V_G \left (n_{1}+n_{0}\right) + \ U n_{1} n_{0} 
-\sum_{p=-\infty}^{\infty} t_{p,p-1} 
(c^\dagger_p c^{\phantom{\dagger}}_{p-1} + h.c.). 
\label{hamiltonian}
\end{equation}
Outside the nano-system, the energy scale is set by a uniform 
hopping amplitude $t_{p,p-1}=1$. Inside the nano-system, the 
hopping amplitude $t_{1,0}=t_d$ is one of the nano-system parameters. 
$c^{\phantom{\dagger}}_p$ ($c^\dagger_p$) is the annihilation (creation) 
operator at site $p$, and $n_p = c^\dagger_p c^{\phantom{\dagger}}_p$. 
In the HF approximation, the ground state is assumed to be a Slater 
determinant of one-body wave-functions $\psi_{\alpha}(p)$ of energy 
$E_{\alpha} < E_F$, $E_F=-2 \cos k_{F}$ being the Fermi energy. 
The effective HF Hamiltonian corresponds to two central sites without 
nearest neighbor repulsion, with renormalized potentials $V_0$ and $V_1$ 
(instead of $V_G$) and hopping amplitude $v$ (instead of $t_d$), coupled 
to two semi-infinite leads. Denoting $\langle c^\dagger_n 
c^{\phantom{\dagger}}_{m} \rangle = \sum_{E_{\alpha}<E_{F}} 
\psi^*_{\alpha}(n)\psi_{\alpha}(m)$, the HF parameters $v,V_0$ and $V_1$ 
are given by the three coupled equations  
\begin{eqnarray}
v=& t_d+U\left \langle c^\dagger_0 c^{\phantom{\dagger}}_{1} (v,V_0,V_1)
\right \rangle 
\label{hf0}
\\
V_0=& V_G+U\left\langle c^\dagger_1 c^{\phantom{\dagger}}_{1} (v,V_0,V_1) 
\right \rangle 
\label{hf1}
\\
V_1=& V_G+U\left\langle c^\dagger_0 c^{\phantom{\dagger}}_{0} (v,V_0,V_1)
\right \rangle,  
\label{hf2}
\end{eqnarray}
which have to be solved self-consistently. 

 Let us first consider the case without the AB-detector. The same set-up 
has been studied \cite{vwj} using the DMRG algorithm, which is valid even 
if $U$ is large (Coulomb blockade without potential barriers). 
There is reflection symmetry, $\langle c^\dagger_1 c^{\phantom{\dagger}}_{1} 
\rangle =\langle c^\dagger_0 c^{\phantom{\dagger}}_{0} \rangle$, 
Eqs.\ (\ref{hf1}) and (\ref{hf2}) are identical, and $V_0=V_1=V$.  
Once the self-consistent values of $v$ and $V$ are obtained, the 
effective transmission coefficient $t_{s}$ reads:  
\begin{equation}
t_{s}=\frac{v \left(1-\exp (-2ik_{F})\right)}
{v^2-\exp (-2ik_{F})-2V \exp (-ik_{F}) - V^2}.
\label{transmission}
\end{equation} 
For this toy model, the analytical form of 
$\langle c^\dagger_p c^{\phantom{\dagger}}_{p'} \rangle$ 
of Eqs.~(\ref{hf0},\ref{hf1},\ref{hf2}) can be given as a function of 
$v$, $V$ and $k_{F}$, as in Ref.\ \cite{afp}. The self-consistent 
values of $v$ and $V$ can be obtained analytically if $U$ is small or 
numerically otherwise, solving the coupled Eqs.\ (\ref{hf0},\ref{hf1}). 
Alternatively, one can diagonalize the HF Hamiltonian numerically 
for leads of finite size $N_L$, make the extrapolation to the limit 
$N_L \rightarrow \infty$, and numerically determine the self consistent 
solution of Eqs.\ (\ref{hf0},\ref{hf1}). 

Moreover, the HF-equations have a simple solution, if one 
makes an approximation which becomes valid when $t_d \gg 1$. 
We only explain the idea in this letter, the detailed calculations will be  
given in a forthcoming paper. The nano-system without leads 
and interaction has only two states of energy $V_G \pm t_d$, separated by 
an energy gap $2t_d$. Let us define the operators $d^{\phantom{\dagger}}_s=
(c^{\phantom{\dagger}}_0+c^{\phantom{\dagger}}_1)/\sqrt2$ and 
$d^{\phantom{\dagger}}_a=
(c^{\phantom{\dagger}}_0-c^{\phantom{\dagger}}_{1})/\sqrt2$, 
$n_s=d^{\dagger}_s d^{\phantom{\dagger}}_s$ and $n_a=d^{\dagger}_a 
d^{\phantom{\dagger}}_a$. When $t_d$ is large and for the values of $V_G$ 
where $|t_s|^2 \neq 0$, the symmetric state of energy $V_G-t_d$ is below 
$E_F$, while the anti-symmetric state of energy $V_G+t_d$ is above $E_F$. 
In that case, the symmetric state is occupied ($\langle n_s \rangle \approx 
1$), the anti-symmetric one is empty ($\langle n_a \rangle \approx 0$), and 
the solution of the HF-equations becomes straightforward. One finds 
$v \approx t_d+U/2$, $V \approx V_G+U/2$, and 
\begin{equation}
|t_s|^2 \approx  \Delta \left ( \frac{\Gamma^2}{(V_G-V_A)^2 - \Gamma^2} - 
\frac{\Gamma^2}{(V_G-V_B)^2 - \Gamma^2} \right),
\label{transmissionHF}
\end{equation}
where $\Delta=(2t_d+U)/(2V_G+U+2 \cos k_F)$, $\Gamma=\sin k_F$, 
$V_A=t_d- \cos k_F$ and $V_B= -t_d -\cos k_F -U$.
%
%
\begin{figure}
\centerline{
\epsfxsize=8.5cm 
\epsfysize=9cm 
\epsffile{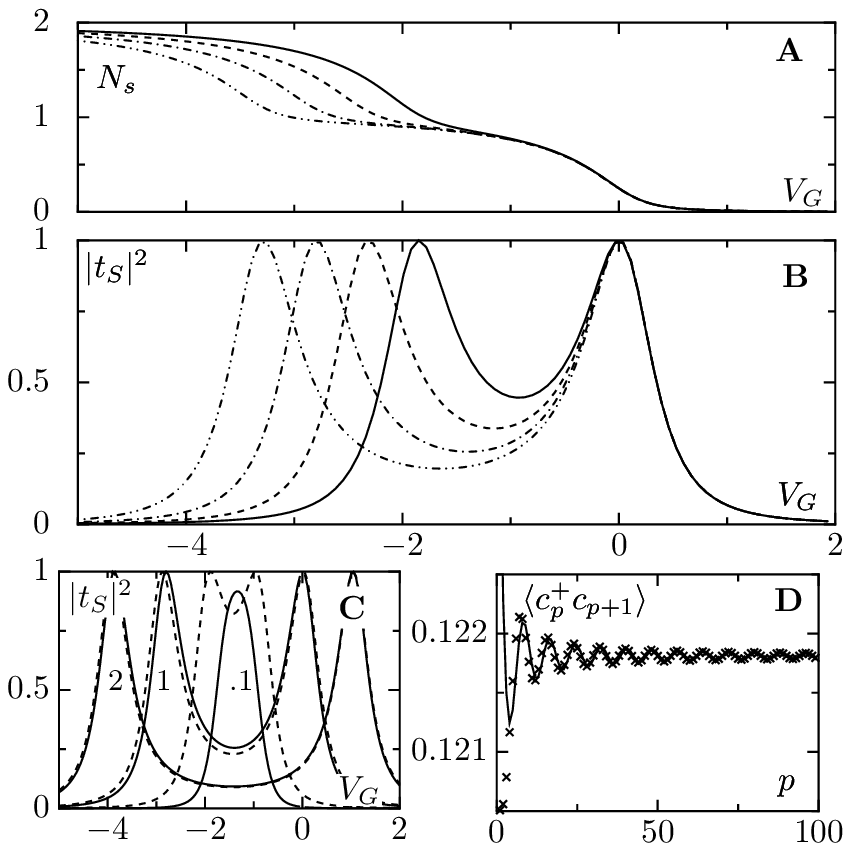}
}
\caption
{
Nano-system without AB-detector for $k_{F}=\pi/8$ (filling $1/8$).  
A: Number $N_s$ of electrons inside the nano-system for $t_d=1$ 
as a function of $V_G$ for $U=0$ (solid), $0.5$ (dashed), $1$ 
(dot-dashed) and $1.5$ (dot-dot-dashed). B: Corresponding transmission 
$|t_{s}|^2$. C: Transmission $|t_{s}|^2$ as a function of $V_G$ for $U=1$ 
and values ($2$, $1$ and $0.1$) of $t_d$ given in the figure: exact HF 
behaviors (solid lines) compared to the behaviors given by 
Eq. (\ref{transmissionHF}) (dashed lines). 
D: $\langle c^{\dagger}_p c_{p+1} \rangle$ as a function of $p$ for $U=1$ 
and $V_G=0$. The solid line is an asymptotic fit 
$F_{a,b,c}(p)$ (Eq.\ (\ref{friedel})) of the HF values ({\sf x}), with  
$a=0.1218$, $b=0.00245$ and $c=-0.38$. $\langle c^{\dagger}_p c_{p} 
\rangle$ (not shown) can be fitted by $F_{a,b,c}(p)$ with $a=1/8$, 
$b=0.0025$ and $c=-0.45$.
}
\label{fig2} 
\end{figure} 
  In FIG.\ \ref{fig2} (B), the nano-system transmission $|t_s|^2$  
without the AB-detector ($t_s$ obtained from Eq.\ (\ref{transmission})), 
is shown as a function of $V_G$, for 4 values of $U$,  $t_d=1$, and a 
Fermi momentum $k_{F}=\pi/8$ (filling $1/8$). The 
corresponding number of electrons inside the nano-system 
$N_s=\langle n_s+n_a \rangle$ is shown above (FIG.\ \ref{fig2} (A)). One 
can see that $|t_s|^2$ exhibits two peaks when $V_G$ decreases, separated 
by an interval $\approx 2t_d+U$, as predicted by Eq. (\ref{transmissionHF}). 
When $t_d=1$, Eq. (\ref{transmissionHF}) does not give an accurate 
description of the effect of $V_G$, this description becoming accurate when 
$t_d$ is larger (see FIG.\ \ref{fig2} (C)). $|t_s|^2 \approx 0$ when the 
nano-system is either empty (large positive $V_G$) or full (large negative 
$V_G$). Decreasing $V_G$, one has a first transmission peak when the 
symmetric state becomes occupied ($V_G \approx V_A$), followed by 
a second peak when the anti-symmetric state is filled ($V_G \approx V_B$). 
When $t_d <1$, the two states become almost degenerate, they are occupied 
at almost the same gate voltage and the two peaks merge to form a single 
peak structure which is not described by Eq. (\ref{transmissionHF}), as 
shown in FIG.\ \ref{fig2} (C). 

 The effect of the nano-system parameters $U$ and $V_G$ upon the leads 
decays as $\langle c^\dagger_p c^{\phantom{\dagger}}_{p'} 
\rangle$ ($p'=p$ or $p+1$) towards their asymptotic values when 
$p \rightarrow \infty$. One finds the usual decay 
\begin{equation}
F_{a,b,c}(p) = a + \frac{b(U,V_G) 
\cos \left( 2 k_{F} p+c(U,V_G) \right)}{p}
\label{friedel}
\end{equation}
of Friedel oscillations inside a 1d non interacting electron gas. 
This is shown in FIG.\ \ref{fig2} (D) for $p'=p+1$. As for 
$p'=p$, these decays are characterized by an 
asymptotic value $a$, an amplitude $b(U,V_G)$ and a phase shift $c(U,V_G)$. 
If one puts a one body scatterer in series with the nano-system, one can 
induce Friedel oscillations inside the nano-system. This will give values 
for the HF parameters $v$, $V_0$, $V_1\neq V_0$ different from 
their values $v$, $V_0=V_1=V$ without the second scatterer. Using as  
second scatterer an AB-detector with a ring, periodic AB-oscillations of 
$v$, $V_0$ and $V_1$ will be induced when a flux is varied through the ring. 

%
%

  The AB-detector sketched in FIG.\ \ref{fig1} includes two 3-lead 
contacts (3LC), the first for attaching the vertical lead to the 
horizontal lead, the second for attaching the ring to the vertical 
lead. A 3LC is made of 4 sites indicated by black circles. Its 
Hamiltonian is given by 
$H_P=-\sum_{p=1}^3 t_{P,p} (c^\dagger_P c^{\phantom{\dagger}}_{p} 
+ h.c.)$, where $t_{P,p}=1$, $P$ denoting the central site and $p$ 
its 3 neighbors. $L_c$ ($L'_{c}$) is the number of sites between 
the upper 3LC and the nano-system (the lower 3LC). $L_R$ is the number 
of sites of the ring, without those of the lower 3LC. A $3 \times 3$ 
matrix $S_P(k)$ describes the scattering by a 3LC at $E=-2 \cos k$. 
$S_P(k)$ has identical diagonal elements $s_{pp}=- e^{ik}/d(k)$ and 
identical off-diagonal elements $s_{pp'}=(2i \sin k)/d(k)= s_{p'p}$ 
where $d(k)=3 e^{ik}-2 \cos k$. 

The reflection amplitude of the ring (vertical lead) threaded by a flux 
$\Phi$ ($\varphi = 2\pi\Phi/\Phi_0$, $\Phi_0$ being the flux quantum) 
reads 
\begin{equation}
r_R(\varphi)=\frac{h_k(\varphi) - \sin (kL_R)}{-h_k(\varphi)+e^{2ik} 
\sin (kL_R)},
\label{boucle}
\end{equation}
where $h_k(\varphi)=2 e^{ik} (\cos (kL_R)- \cos \varphi) \sin k$. 
The reflection and transmission amplitudes of the AB-detector 
(in the horizontal lead) read
\begin{equation}
r_{AB}(k)=\frac{-e^{2ik}-e^{2ikL'_{c}} r_R(\varphi)}
{2 e^{2ik}-1+r_R(\varphi) e^{2ik(L'_c+1)}}
\label{r_{AB}}
\end{equation}
\begin{equation}
t_{AB}(k)=\frac{2i \sin k e^{ik} (1+e^{2ikL'_{c}}r_R(\varphi))} 
{2e^{2ik}-1 +r_R(\varphi)e^{2ik(L'_c+1)}}
\label{r_R}.
\end{equation}
\begin{figure}
\centerline{
\epsfxsize=8.5cm 
\epsfysize=8cm 
\epsffile{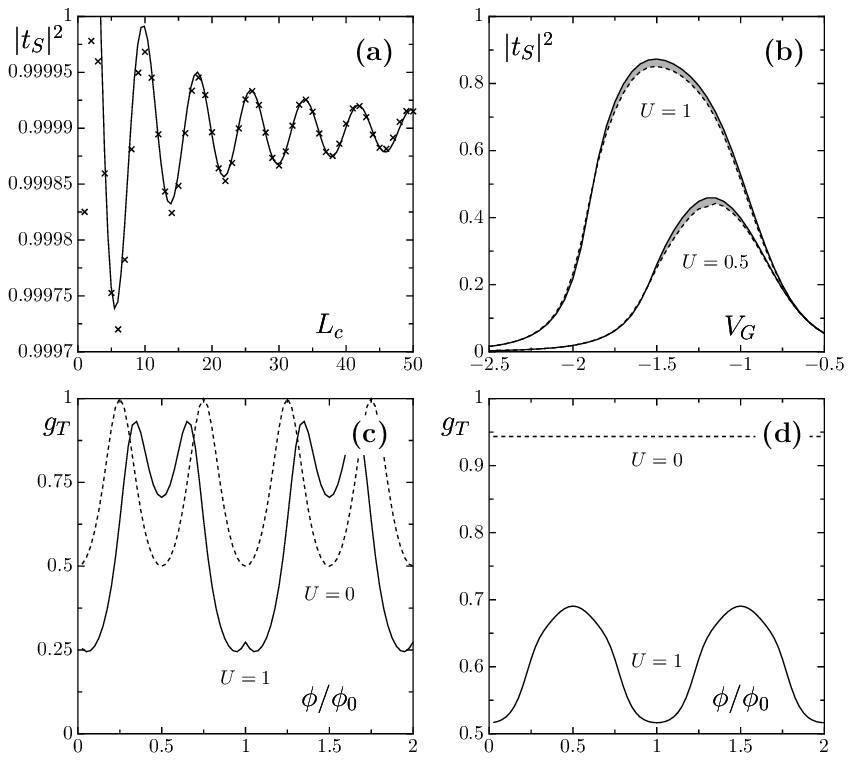}
}
\caption
{Nano-system with AB-detector for $L'_c=4$. 
A: Transmission $|t_s|^2$ as a function of $L_c$ for $t_d=1$, $U=1$, 
$k_F=\pi/8$, $L_R=7$, $V_G=-2.8$ and $\Phi=0$. HF values ({\sf x}) and 
fit $F_{a,b,c}(p)$ (Eq.\ (\ref{friedel}) - solid line) with $a=0.9999$, 
$b=0.00092$, $c=1.678$. B: $|t_s|^2$ as a function of $V_G$ for $t_d=0.1$, 
$k_F=\pi/8$, $L_R=7$ and $L_c=2$. The dashed (solid) curves correspond to 
$\Phi= 0$ ($\Phi_0/2$). The grey areas underline the effect of $\Phi$. 
C: Conductance $g_T$ of the nano-system and the AB-detector in series, 
as a function of $\Phi/\Phi_0$ for $t_d=0.1$ and $k_F=\pi/2$. $L_c=2$, 
$L_R=7$ $V_G=-0.8$. $U=1$ (solid line) and $U=0$ (dashed). D: $g_T$ as 
a function of $\Phi/\Phi_0$ for $L_c=2$ and $L_R=4$ ($\sin k_FL_R=0$). 
$t_d=0.1$, $k_F=\pi/2$ and $V_g=-0.7$. $U=1$ (solid) and $U=0$ (dashed).
}
\label{fig3} 
\end{figure} 
 We now study the nano-system in series with the AB-detector, solving 
numerically the Eqs.\ (\ref{hf0}-\ref{hf2}) for having the values 
$v(\varphi,L_c), V_0(\varphi,L_c)$ and $V_1(\varphi,L_c)$ characterizing 
the nano-system. FIG.\ \ref{fig3} (A) shows the effect of the AB-detector 
upon the nano-system transmission $|t_s(L_c)|^2$, as a function of 
$L_c$. $t_s(L_c)$ is given by extending formula (\ref{transmission}) to 
the case where $V_0 \neq V_1$. The transmission $|t_s(L_c)|^2$ exhibits 
decaying oscillations towards the asymptotic value characterizing the 
nano-system without AB-detector. The decay is given by a function 
$F_{a,b,c}(L_c)$. 

 For $t_d=1$, the effect of the AB-detector upon $|t_s|^2$  remains 
negligible ($\approx 10^{-4}$), even for small values of $L_c$. 
This effect can be made $10^3$ times larger if $t_d$ is reduced by a 
factor $10$. We have shown that $|t_s|^2$ is given by Eq. 
(\ref{transmissionHF}) when $t_d > 1$, in the limit where 
$\left \langle n_s \right \rangle \approx 1$ and $\left \langle n_a \right 
\rangle \approx 0$. In this limit, one cannot strongly vary 
$\left \langle n_s \right \rangle$ and $\left \langle n_a \right \rangle$ 
by the Friedel oscillations of the AB-detector, and the HF parameters are 
almost independent of $\varphi$. But much larger AB-oscillations of the HF 
parameters become possible if $t_d < 1$, when two almost degenerate levels 
are near $E_F$ for the same value of $V_G$. This is shown in FIG.\ \ref{fig3} 
(B) for $t_d = 0.1$. The two 
transmission peaks shrink to form a single peak structure which depends on 
the flux $\varphi$ threading the ring. In FIG.\ \ref{fig3} (B), putting 
$\Phi_0/2$ through the ring increases $|t_s|^2$ by a visible amount (grey 
areas), an effect $100$ times larger than when $t_d=1$. To make the effect 
even larger, one can adjust the wave-length of the Friedel oscillations to 
the size of the nano-system. Increasing $k_F$ from $\pi/8$ to $\pi/2$ 
(half-filling), the size of the AB-oscillations of $|t_s|^2$ can be 
increased by another factor $\approx 10$. 

  Having reduced the many-body scatterer to an effective one body scatterer, 
the conductance is given by the Landauer formula valid for a bare one 
body scatterer. Let us consider the two probe geometry of FIG.\ \ref{fig1}, 
and study the conductance $g_T$ of the nano-system and the AB-detector in 
series. $g_T=|t_T|^2$ (in units of $e^2/h$), where $t_T$ is given by the 
combination law: 
\begin{equation}
t_T=t_s \frac{e^{ik_{F}L_c}}{1- r'_s r_{AB} e^{2ik_{F}L_c}} t_{AB} .
\end{equation}
$r'_s$ ($r_{AB}$) is the reflection amplitude of the nano-system 
(of the AB-detector). Because $r_{AB}$ and $t_{AB}$ 
depend on $\varphi$, $g_T$ exhibits AB-oscillations even without 
interaction, when $t_s$ and $r'_s$ are independent of $\varphi$ ($U=0$ 
or $L_c$ too large). However, when the electrons interact inside the 
nano-system and $L_c$ is not too large, $t_s$ and $r'_s$ exhibit also 
AB-oscillations around certain values of $V_G$, which can strongly modify 
the AB-oscillations of $g_T$. In FIG.\ \ref{fig3} (C), one can see how 
the shapes of the AB-oscillations are modified by the interaction, while 
their amplitudes are increased. FIG.\ \ref{fig3} (D) corresponds to a 
case where the ring is perfectly reflecting at $E_F$ (Eq. (\ref{boucle}), 
$r_R=-1$ when $\sin (k_{F} L_R)=0$). In that special case, $t_{AB}$ and 
$r_{AB}$ are independent of $\varphi$ at $E_F$. But $g_T$ does have 
AB-oscillations when the electrons interact inside the nano-system, the 
HF corrections depending on HF states below $E_F$ for which 
$\sin (k_{\alpha}L_R) \neq 0$. In this special case, the very large 
AB-oscillations of $g_T$ are a pure many body effect.

  The effect of the flux upon the HF parameters can be also detected if one 
uses 4 probes instead of 2. The idea \cite{engquist} is to weakly 
contact 2 additional probes near the nano-system, for measuring 
directly the voltage drop at its extremities, and not on a larger scale 
including the AB-detector. However, since our effect requires to have 
the nano-system and the AB-detector inside the same quantum coherent region, 
the obtained conductance is no longer given by the 2 probe formula 
$g_s=|t_s|^2$, but by the multi-terminal formula \cite{buettiker} derived 
by B\"uttiker. In a 4 probe set-up, this formula yields also non local 
effects without interaction, which have been observed in mesoscopic 
conductors, using metallic wires \cite{washburn} or semi-conductor 
nanostructures \cite{skocpol}, where the interaction is too weak for 
making our many body effect important. But the non local effect seen in 
Refs. \cite{washburn,skocpol} should be strongly enhanced, if a region 
where the electrons interact is included between the 2 voltage probes.  

%
%

 We have studied spinless fermions (polarized electrons) and 1d leads, 
and shown that an attached ring can considerably modify the nano-system 
transmission for well chosen values of $U$, $V_G$, $t_d$ when $L_c$ is 
not too large. The HF approximation could be easily extended to higher 
dimensions. To include the spins could be more difficult. The double 
occupancy of each site becoming possible, a Hubbard repulsion must be added, 
making our double site model slightly more complicated than the Anderson 
model used for the Kondo problem \cite{hewson}. The study of the effect of 
flux dependent Friedel oscillations upon the Kondo problem is left for future 
investigations. 

%
%

 We thank G.~Faini, M.~Sanquer and specially D.~Weinmann for useful comments. 
The support of the network ``Fundamentals of nanoelectronics'' of the EU 
(contract MCRTN-CT-2003-504574) is gratefully acknowledged.

\end{document}